\documentclass[10pt]{iopart}

\expandafter\let\csname equation*\endcsname\relax

\expandafter\let\csname endequation*\endcsname\relax
\usepackage{amsmath}

\usepackage{caption}
\usepackage{cite}
\usepackage{amssymb,amsfonts}
\usepackage{algorithmic}
\usepackage{graphicx}
\usepackage{textcomp}
\usepackage{comment}
\usepackage{xcolor}

\definecolor{mango}{rgb}{1.0, 0.51, 0.26}
\definecolor{blue-violet}{rgb}{0.54, 0.17, 0.89}

\newcommand{\rhoRev}[1]{\textcolor{black}{#1}}

\newcommand{\fb}[1]{\textcolor{black}{#1}}
\newcommand{\fbagain}[1]{\textcolor{black}{#1}}
\newcommand{\rhoEdit}[1]{\textcolor{black}{#1}}

\newcommand{\gian}[1]{\textcolor{black}{#1}}
\newcommand{\fbRev}[1]{\textcolor{black}{#1}}

\newcommand{\RG}[1]{\textcolor{black}{#1}}

\begin{document}

\title[Multimodal Integration of Olfactory and Visual Processing through DCM analysis]{Multimodal Integration of Olfactory and Visual Processing through DCM analysis: Contextual Modulation of Facial Perception}

\author{Gianluca Rho$^{1,2,*}$, Alejandro Luis Callara$^{1,2}$, Francesco Bossi$^{1}$, Dimitri Ognibene$^{3,4}$,  Cinzia Cecchetto$^5$, Tommaso Lomonaco$^6$, Enzo Pasquale Scilingo$^{1,2,\dagger}$, and Alberto Greco$^{1,2,\dagger}$}

\address{$^1$Dipartimento di Ingegneria dell'Informazione, University of Pisa, Pisa, Italy}
\address{$^2$Research Center “E. Piaggio”, School of Engineering, University of Pisa, Pisa, Italy}
\address{$^3$Università Milano-Bicocca, Milan, Italy}
\address{$^4$University of Essex, Colchester, UK}
\address{$^5$Department of General Psychology, University of Padova, Italy}
\address{$^6$Department of Chemistry and Industrial Chemistry, University of Pisa, Italy}
\address{$^\dagger$These authors contributed equally to the work}
\ead{$^*$gianluca.rho@phd.unipi.it}
\vspace{10pt}

\begin{abstract}
This study examines the modulatory effect of contextual hedonic olfactory stimuli on the visual processing of neutral faces using event-related potentials (ERPs) and effective connectivity analysis. The aim is to investigate how odors' valence influences the cortical connectivity underlying face processing, and the role \gian{arousal enhanced by} faces plays on such visual-odor multimodal integration.
To this goal, a novel methodological approach combining electrodermal activity (EDA) and dynamic causal modeling (DCM) was proposed to examine cortico-cortical interactions \gian{changes}.
\gian{The results revealed that EDA sympathetic responses were associated with an increase of the N170 amplitude, which may be suggested as a marker of heightened arousal to faces.}
Hedonic odors had an impact on early visual ERP components, with increased N1 amplitude during the administration of unpleasant odor and decreased vertex positive potential (VPP) amplitude during the administration of both unpleasant and neutral odors.
\gian{On the connectivity side}, unpleasant odors strengthened the forward connection from the inferior temporal gyrus (ITG) to the middle temporal gyrus (MTG), involved in processing changeable facial features. 
Conversely, \gian{the occurrence of} sympathetic responses \gian{was correlated with an inhibition of} \gian{the same} connection, and \gian{with an} enhance\gian{ment of} the backward connection from ITG to the fusiform face \gian{gyrus.}
These findings suggest that negative odors \gian{may} enhance the interpretation of emotional expressions and mental states, \gian{while faces capable of enhancing} sympathetic arousal prioritize the processing of identity.
The proposed methodology provides insights into the neural mechanisms underlying the integration of visual and olfactory stimuli in face processing.
\end{abstract}

%
\ioptwocol

\section{Introduction}
\label{sec:introduction}

The processing of facial expressions is a fundamental mechanism for perceiving others' intentions and emotions \cite{syrjanen2021review}. In real-life situations, this process is not limited to the visual system alone but is influenced by contextual information from other sensory channels \cite{aviezer2017inherently}. Olfactory stimuli have been found to play a crucial role in modulating the hedonic perception of faces  \cite{syrjanen2021review, damon2021olfaction}. Previous studies have shown that the emotional valence conveyed by contextual odors can affect the recognition of facial expressions and subjective ratings of faces \cite{li2020unpleasant, cook2015pleasant, cook2017pleasant, cook2018simultaneous, syrjanen2018background, syrjanen2019background}.

These behavioral responses are accompanied by physiological changes, as evidenced by electroencephalographic (EEG) event-related potential (ERP) studies. These studies have \gian{reported} an effect of \gian{the valence of the }odors at both early sensory stages (P1/N1, N170, and Vertex Positivity Potential (VPP)) \cite{syrjanen2018background, adolph2013context, leleu2015contextual, steinberg2012rapid} and later cognitive stages (late positive potential) of face processing \cite{callara2022human, hartigan2017disgust, rubin2012second}.
\gian{However, ERP waveforms reflect the overall activation of the face-perception system, an ensemble of interconnected regions which categorize visual stimuli as faces by analyzing various factors such as expression, gender, and identity \cite{rossion2014understanding}.}
The close association between olfactory and visual areas has been established in previous research \cite{zatorre2000neural, jadauji2012modulation}. 
Therefore, \gian{investigating the effect of emotional odors on the interactions among the areas of the face-perception system} could provide valuable insights into the integration mechanism between olfaction and vision. Such investigations can enhance our understanding of how sensory cues interact to shape our perception of others' facial expressions and emotional states.

\gian{T}he standard ERP analysis based on components' amplitude and latency can not provide \rhoEdit{a sufficient level of detail to investigate brain dynamics at the network level.}
Nevertheless, ERPs can be combined with more sophisticated techniques to provide a window on the cortical sources' dynamics underlying face processing. 
In this context, EEG connectivity analysis allows to investigate the interaction among neuronal assemblies \cite{friston2011functional}. Particularly, effective connectivity \RG{estimates} the direct and directional (i.e., causal) influence that a source exerts over another \cite{friston2011functional}. 
Among the various \rhoEdit{effective connectivity} approaches \cite{schoffelen2009source,friston2011functional}, Dynamic Causal Modeling (DCM) is a powerful model-based technique designed to test for the effect\fb{s} of experimental factors on the interactions among regions of a specified brain network,  starting from observed electrophysiological or functional imaging responses \cite{friston2003dynamic, kiebel2008dynamic}.  
Several studies applied DCM to reveal the dynamics among sources of the face-perception network either through fMRI \cite{frassle2016handedness, nguyen2014fusing, kessler2021revisiting, li2010effective, fairhall2007effective}, MEG \cite{chen2009forward, garvert2014subcortical}, or intracranial EEG \cite{sato2017bidirectional} recordings. 
\RG{In this light, DCM can be combined with the information provided by visual ERPs to investigate the modulatory effect of hedonic contextual odors on the cortico-cortical interactions among face-processing areas.}

A \gian{potential factor of interest} in the analysis of the face-odor multimodal integration concerns the \gian{arousal elicited by} faces.
Specifically, within an event-related paradigm, several repetitions of the stimulus are presented over time, and
it could be hypothesized that not
all the faces are able to elicit \gian{the same affective} response due to the \gian{subjective saliency of facial features \cite{lang1998emotional, balconi2009arousal, eimer2000face}}.
This may affect the signal-to-noise ratio (SNR) of the observed ERPs, and potentially reduce or even obscure the modulatory effect of contextual odors on the connectivity. \RG{However, although \gian{arousal} has \gian{already} been shown to influence \gian{face-evoked} ERPs \cite{almeida2016perceived, hietanen2014additive}, its effects when faces are presented with concomitant olfactory stimuli are still poorly \gian{investigated}.}
Arousal has a crucial effect on the processing of external inputs. Motivational\fb{ly} relevant stimuli, such as the perception of a face \fb{with specific \gian{intrinsic} features}, are able to generate a \gian{transitory} and automatic enhancement of arousal \cite{lang1995emotion, lang1998emotion}\RG{,} that \gian{entails a prioritized processing of the stimulus in the visual stream \cite{hietanen2014additive}}.
This mechanism is associated with a series of physiological changes, including \gian{a top-down (i.e., endogenous) affective influence on sensory gain control \cite{hietanen2014additive}, an increase in the amplitude of frequency-specific brain oscillations \cite{balconi2008consciousness,balconi2009arousal} and in the magnitude of ERP components \cite{almeida2016perceived, hietanen2014additive}.}
\fb{In this light, testing for the actual occurrence of \gian{enhanced arousal} is crucial to} ensure that a face has been successfully perceived, thus maximizing the SNR of ERP responses and the consequent effect size of odors’ modulation.

Modeling the effect\fb{s} of \gian{arousal} \RG{at both the ERP and DCM connectivity levels} requires knowing in advance whether a stimulus has the property of eliciting \gian{it}. However, such a prior knowledge is far from being easily \fb{identified}, since particular features that could characterize a perceived stimulus as relevant (e.g., facial identity, eye gaze, expression) may vary on a subjective basis. 
\gian{Besides brain activity, arousal is known to influence autonomic nervous system (ANS) dynamics. Particularly, states of high arousal are associated with an increase of peripheral sympathetic activity \cite{boucsein2012electrodermal}. In this light, }
electrodermal activity (EDA) can be exploited as an objective means to identify the occurrence of \gian{enhanced arousal} to a given stimulus\gian{, possibly related to intrinsic facial features or to emotional expressions} \cite{nieuwenhuis2011anatomical, frith1983skin, barry1993stimulus, spinks1985role}.
EDA is comprised of a slow-varying tonic component overimposed to a fast-varying phasic component. Particularly, the latter \fb{is represented by} a series of stimulus-evoked skin conductance responses (SCRs), whose elicitation is driven by peripheral sympathetic neural bursts: i.e., the sudomotor nerve activity (SMNA) \cite{boucsein2012electrodermal}.  
Accordingly, sympathetic responses to faces observed from SMNA can be adopted as a reliable marker of enhanced arousal.  

\rhoEdit{In this work, we investigate the effects of hedonic olfactory stimuli and \gian{arousal} enhancement on face perception, as measured by ERP \fb{components} and source effective connectivity. Specifically, we hypothesize that odors' valence modulates the strength of specific pathways within the face-perception network, and that \gian{arousal evoked by the saliency} of faces could play a role on such a multimodal sensory integration.
To this aim, we acquired the EDA and EEG signals from 22 healthy volunteers performing a passive visual stimulation task with neutral faces and background pleasant, neutral and unpleasant odors.}
We propose a novel methodological approach based on the convex-optimization-based EDA (cvxEDA) framework \cite{greco2015cvxeda} to identify face-evoked peripheral sympathetic responses and characterize the stimuli according to their property of eliciting arousal. 
The outcome of this classification procedure is then adopted to model the effect\fb{s} of odors' valence \gian{and arousal} as between-trial factors on the visual ERP \fb{component}s evoked by faces. We then exploit such ERPs to carry out a DCM and Parametric Empirical Bayes (PEB) analysis of odors and \gian{arousal} on the connectivity among cortical sources found to be activated by the task. \rhoEdit{Particularly, PEB allows to build a hierarchical analysis of effective connectivity, where single-subject estimates about neuronal parameters of interest (e.g., connections' strength) are treated as \gian{stochastic} effects at the group level \cite{zeidman_guide_2019}.}
To validate our aforementioned hypothesis on the role of enhanced \gian{arousal} to faces, we aim to asses\fb{s}: \fb{(i)} whether such mechanism is effectively reflected by the observed ERPs, and \fb{(ii)} \gian{the modulation operated by odors on cortico-cortical interactions at different levels of \gian{arousal}.}

\section{Material and Methods}

\subsection{Participants}

Twenty healthy volunteers (age 27 $\pm$ 3, 5 females) were enrolled in the study. 
\gian{Volunteers did not report any history of neurological and cardiovascular diseases, anosmia or being tested positive to COVID-19 over the past 6 months.}
Volunteers were not allowed to have any food nor drink in the 30 minutes preceding the experiment. 
The study was conducted according with the guidelines of the Declaration of Helsinki\RG{, and approved by the Bioethics Committee of the University of Pisa Review No. 14/2019, May 3rd, 2019}.

\subsection{Olfactory stimuli}
We selected three different odorants: i.e., banana (isoamyl acetate; $CH_3COOCH_2CH_2CH(CH_3)_2$), isovaleric acid ($(CH_3)_2CHCH_2COOH$), and n-butanol ($CH_3CH_2CH_2CH_2OH$). We chose such odorants to convey positive (banana), negative (isovaleric acid), and neutral (n-butanol) valence according to previous literature results\cite{hummel1997sniffin, naudin2012state}. For each subject, we prepared $1ml$ isointense solutions diluting pure odorant substances in distilled water according to the ratios of 1/20 (banana, n-butanol) and 1/10 (isovaleric acid) respectively. Odorants were delivered through an in-house built computer-controlled olfactometer at a flow rate of $60ml/min$.

\subsection{Olfactometer device}
The 4-channels computer-controlled olfactometer used herein is composed by i) a pressure regulator to set air pressure at 7Bar, ii) four stainless-steel containers (50mL) equipped with o-rings, stainless steel caps and clamping rings to ensure a gas-tight closure, iii) 10 low dead-volume three-way solenoid valves (Parker Hannifin, Italy), iv) a digital flow meter (Honeywell, Italy), and v) a disposable nasal cannula. An Arduino in-house code controlled the solenoid valves allowing them being opened and closed through a well-defined sequence of actions. The software allows the delivery of pure clean air (i.e., all valves are opened) or odors kept at room temperature within the containers. Components were connected to each other using polytetrafluoroethylene (PTFE) fittings and tubings (internal diameter of 0.3mm) to reduce the olfactometer dead-volume up to 1mL. The nasal cannula is connected to the olfactometer though a 3 meter long PTFE line. The olfactometer showed a negligible memory effect and a low background emissions of chemicals in the air/odors mainstream as determined by thermal desorption coupled to gas-chromatography and mass spectrometry protocol \cite{lomonaco2021stability}, an overall air flow delivery variability less than 1$\%$, and a rise time (10-90$\%$ of the final value) close to 300 ms.

\subsection{Visual stimuli}
For the visual stimuli of neutral faces, we used the Chicago Face Database \cite{ma2015chicago}. We chose 128 different actors showing a neutral expressions. Particularly, we selected a balanced number of actors across gender, age, and ethnicity, to mitigate potential confounding effects related to the intrinsic characteristics of the actors. The pictures were presented in a completely randomized order with respect to the olfactory condition.
The visual stimuli were shown on a 15'' laptop screen with a refresh rate of 60Hz and a resolution of 1920x1080. The pictures' size was 15cm in width and 10cm in height, and were displayed at about 50cm from the eyes of the participants, resulting in a visual angle of about 17°.

\subsection{Experimental protocol}
\begin{figure*}[h]
	\centerline{\includegraphics[width=0.9\linewidth]{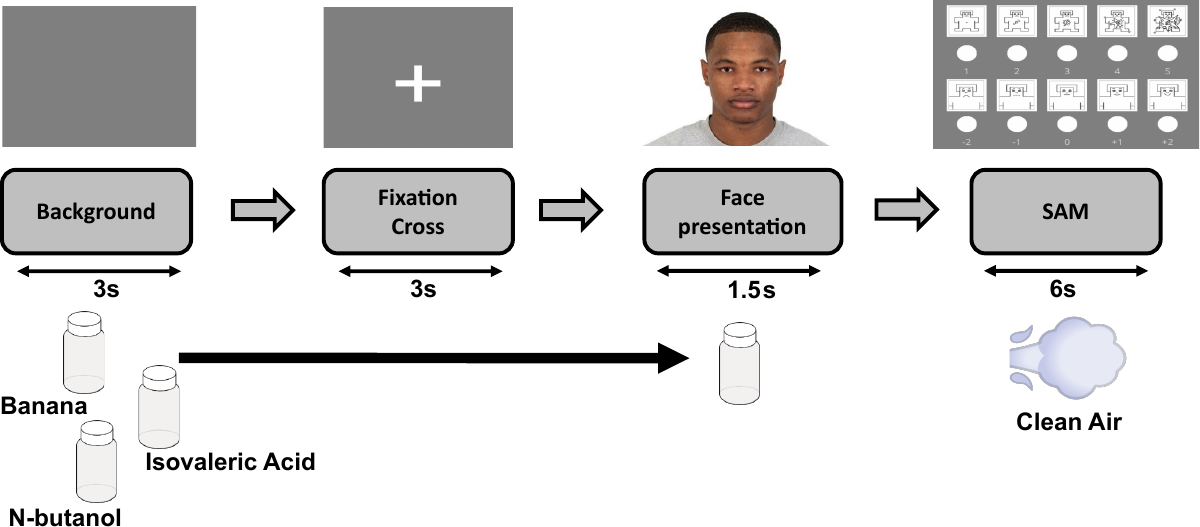}} 
	\caption{Schematic illustration of the experimental protocol. Each trial consisted of: (1) 3s of dark gray background; (2) 3s of fixation cross; (3) 1.5s of neutral face image presentation; (4) 6s of inter-trial rest where subjects rated the valence and arousal of neutral faces according to the self-assessment manikin (SAM) test. A random odor among banana, n-butanol, isovaleric acid and clean air was delivered starting from (1) to the end of (3). A wash-out with clean air was performed during (4).}
	\label{fig:EEG_EDA_expProtocol}
\end{figure*}

The experimental protocol was divided into two parts. First, we presented the three odorants in a randomized order for a duration of about 10s, and we asked the participants to evaluate the perceived hedonic content in terms of valence (from -2 to +2) and arousal (from 1 to 5) according \fb{to} the Self-Assessment Manikin (SAM) test \cite{bradley1994measuring}.
Second, we designed an experimental protocol comprised of 128 trials. \gian{As schematically reported in Fig.\ref{fig:EEG_EDA_expProtocol},} each trial consisted of: (1) 3s of dark gray background; (2) 3s of a dark gray background with a white fixation cross; (3) 1.5s of neutral face image presentation; (4) 6s of inter-trial rest. Within each inter-trial rest, subjects were asked to evaluate the facial expression in terms of valence and arousal according with the SAM test, through an interactive interface. The facial images were presented in combination with clean air or one of the three different odorants scored in the first part of the protocol, for a total of 128/4=32 trials for each olfactory condition. Each odorant was delivered starting from the onset of the dark gray background to the end of the visual stimulus. A wash-out with clean air was performed during the inter-trial rest. 
We presented both visual and olfactory stimuli in a randomized order through the PsychoPy software \cite{peirce2007psychopy}

\subsection{EEG and EDA acquisition}
EEG \fb{signal} was acquired using a high-density 128-channel geodesic EEG System 300 from Electrical Geodesic, Inc. (EGI). Electrodes were grounded through two additional channels placed between Cz and Pz and referenced through Cz. We always kept electrode impedances below 20$k\Omega$ during the acquisitions. EEG was acquired at the sampling frequency of 500Hz.

EDA was acquired using a Shimmer3 GSR+ unit (Shimmer, USA) at the sampling frequency of 250Hz. We recorded EDA through a pair of Ag/AgCl electrodes placed on the proximal phalanx of the first and second fingers of the non-dominant hand, respectively. 

\subsection{EDA-driven sympathetic activity estimation}
We implemented a procedure based on the analysis of EDA signal to estimate the occurrence of enhanced sympathetic responses associated with the visual presentation of faces. 
A well-known problem in EDA analysis concerns the temporal overlapping of consecutive SCRs, which may hamper the association between a given response and its potential triggering stimulus \cite{boucsein2012electrodermal, greco2015cvxeda, greco2016emotions}. 
To address this issue, we adopted the cvxEDA \cite{greco2015cvxeda} \RG{model} to recover an estimate of the SMNA from the observed EDA responses. Specifically, cvxEDA considers that each SCR is preceded in time by sparse and discrete bursts of SMNA. These bursts are characterized by a higher temporal resolution compared to phasic activity, and can thus be exploited to identify the time instants at which peripheral sympathetic responses evoked by faces occur \cite{greco2015cvxeda}.
Accordingly, we assumed that visual stimuli elicit\fb{ing} an enhanced peripheral sympathetic response could be followed by the occurrence of an SCR and, thus, a non-zero SMNA neural burst.  

\begin{figure}[h]
	\centerline{\includegraphics[width=1\linewidth]{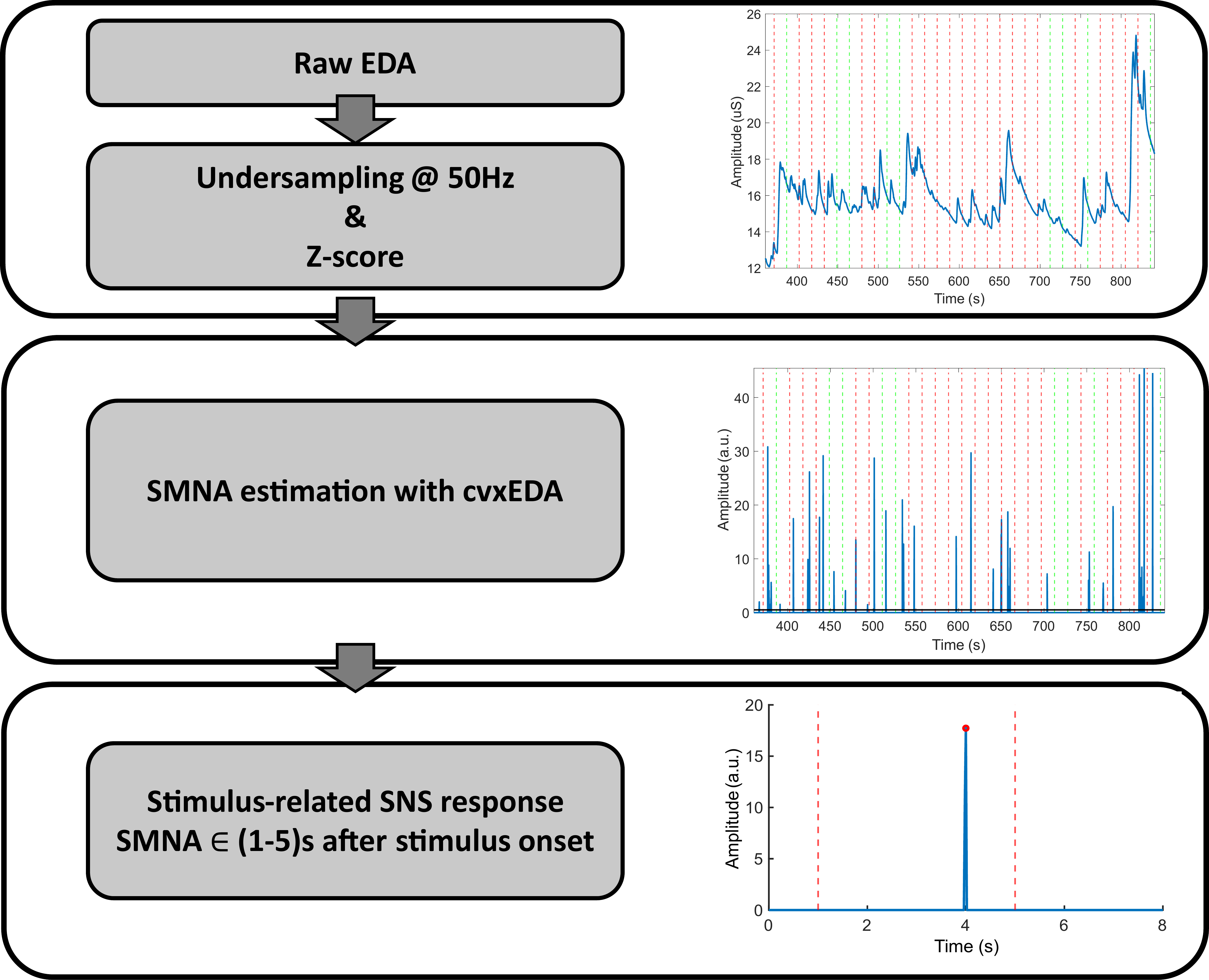}} 
	\caption{Schematic illustration of the proposed procedure to assess stimuli-related peripheral sympathetic responses from the EDA signal. The raw EDA was preprocessed by downsampling at the frequency of 50Hz and applying a Z-score transformation in order to be given as input of the cvxEDA algorithm.
	SMNA estimates are then epoched in the (0-8)s interval with respect to the onset of each stimulus (i.e., 0s).
	Significant sympathetic responses to the stimuli are identified as non-zero SMNA bursts occurring in the (1-5)s interval after stimulus onset}
	\label{fig:EEG_EDA_sympEstimation}
\end{figure}

Operationally, for each subject and for each odor condition (i.e., pleasant, unpleasant, neutral, air), we undersampled the EDA to the sampling frequency of 50Hz, and we performed a Z-scoring on the data \RG{\cite{greco2015cvxeda}}. Then, we applied cvxEDA to obtain an estimate of the SMNA. We identified discrete events of enhanced sympathetic arousal associated with the presentation of visual stimuli as those epochs having non-zero SMNA bursts occurring in the (1-5)s after stimulus onset. This choice is supported by several studies indicating that a stimulus-evoked SCR is observed to occur within that range of latency after stimulus’ onset \cite{boucsein2012electrodermal, sjouwerman2019latency}.
A schematic illustration of the procedure is reported in Fig.\ref{fig:EEG_EDA_sympEstimation}. \\For each odor condition, we then extracted: (1) the number of epochs with/without a stimulus-related sympathetic response (nSymp), (2) the average latency of sympathetic responses, (3) the average amplitude of the SMNA\gian{, computed as the average amplitude across epochs and then over time}, and (4) the subject-average SCR generated by the SMNA bursts.

\subsection{EEG preprocessing}
\label{sec:EEG_prep}
We preprocessed the EEG signal using EEGLAB \cite{delorme2004eeglab}. First, we filtered the data with a zero-phase \gian{low-pass} antialiasing filter and then undersampled it to the sampling frequency of 100Hz. Afterwards, we applied a zero-phase high-pass filter at the cutoff frequency of 0.1Hz to improve data stationarity. We removed flat and
poorly correlated channels by exploiting the method presented in \cite{mullen2015real}. Specifically, each channel was compared with its reconstructed version obtained from the spherical interpolation of its neighbors and was removed if the correlation coefficient was less than a user-defined threshold. Here, we used a correlation threshold of 0.8.
After visual inspection, we recovered the removed channels through spherical interpolation, and we re-referenced the data to its average. For each subject, we epoched the EEG signal from -200ms to 1000ms with respect to the onset of visual stimuli (i.e., 0ms), and we \rhoRev{visually} inspected the data to remove epochs contaminated by artifact activity. \rhoEdit{A number of 125 $\pm$ 7 (mean $\pm$ standard deviation) epochs was retained over the subjects.}
Finally, we decomposed EEG data through ICA \cite{makeig1995independent}, and we removed ICs resembling artifact activity (e.g., muscular, ocular and other sources of noise) \fb{through visual inspection \rhoEdit{of their associated time course, scalp map, and power spectrum}}.
Clean EEG epochs were corrected for their baseline (i.e., from -200ms to 0ms) and low-pass filtered at the cutoff frequency of 30Hz using default settings in EEGLAB. 
We \gian{grouped the epochs according to the odor condition, and we further distinguished them based on the presence or absence of a peripheral sympathetic activation (hereinafter referred to as \textit{Symp} condition). }
Accordingly, we obtained subject-average ERPs for \gian{a total of 8 conditions: clean air, n-butanol, banana, and isovaleric acid, with/without the presence of a sympathetic response.}

\gian{We focused on} relevant ERP components associated with the visual processing of faces, as well as their typical regions of interest (ROIs) on the scalp, \gian{based on the} previous literature \cite{syrjanen2018background, leleu2015contextual, zhang2014spatial, trautmann2013perception, rossion2014understanding, rubin2012second, adolph2013context} \gian{(see Table \ref{tab:ERP_components} for details)}.
\begin{table*}[]
	\centering
	\caption{Channels region of interest (ROI) and time interval (ms) for each of the investigated ERP components. Channels' name is reported according to the geodesic EGI 128-channels cap.}
	\setlength\tabcolsep{3pt}
	\begin{tabular}{llc}
		\hline
		& \multicolumn{1}{c}{Channels}                                                                                      & \multicolumn{1}{l}{Time interval} \\ \hline
		P1 & E59    E65    E66    E70    E71    E76    E83    E84    E90    E91                                                            & 120-160 ms                                \\
		N1 & \begin{tabular}[c]{@{}l@{}}E6     E7    E13    E30    E31    E37    E54    E55    E79    E80    E87\\   E105   E106   E112   E129\end{tabular} & 120-160 ms                                \\
		rN170 & E96    E97   E101   E102   E108                                                                                       & 160-200 ms                                \\
		lN170 & E45    E46    E50    E51    E58                                                                                         & 160-200 ms                                \\
		VPP & \begin{tabular}[c]{@{}l@{}}E7    E13    E30    E31    E37    E54    E55    E79    E80    E87\\   E105   E106   E112   E129\end{tabular} & 160-200 ms                                \\
		P2 & E59    E65    E66    E70    E71    E76    E83    E84    E90    E91                                                                    & 220-300 ms                                \\  
		LPP & E52    E53    E60    E61    E62    E67    E77    E78    E85    E86    E92 & 300-500 ms                                \\ 
	\end{tabular}
	\label{tab:ERP_components}
\end{table*}
Specifically, we extracted: \gian{(1)} the P1 component in the 120-160ms interval after stimulus onset in the occipital region around O1 and O2, \gian{(2)} the N1 component at the same latency as P1, in the central region around Cz, \gian{(3)} the \gian{right/left} N170 component in the 160-200ms interval after stimulus onset in the parieto-temporal regions near P7 and P8, respectively, and \gian{(4)} the Vertex \fb{Positive} Potential (VPP) in the 160-200ms interval after stimulus onset in the same ROI of the N1 component. Moreover, for a comparison with the previous literature, we also identified: \gian{(5) the} P2 component at 220-300ms after stimulus onset in the same ROI of P1, and \gian{(6)} the Late Positive Potential (LPP), a sustained positivity at 300-500ms after stimulus onset around Pz. 
For each subject and for each of these components, we extracted the mean amplitudes across the \gian{respective} time intervals and ROIs.

\subsection{\RG{Statistical analysis on self-report, EDA and ERP data}}

\gian{We investigated for possible differences in the valence and arousal subjective ratings of olfactory stimuli through multiple Wilcoxon sign-rank tests\gian{, with a significance level of $\alpha=0.05$}.}
P-values were adjusted for multiple comparison\fb{s} testing with Bonferroni correction. 
\gian{Analogously, we tested for an effect of olfactory stimuli on the valence and arousal subjective ratings of faces through pairwise Wilcoxon tests ($\alpha=0.05$).}

\RG{Concerning the physiological data, }we tested for significant differences in the amplitude and latency of SMNA responses across odor conditions through separate 1x4 within-subject ANOVAs \gian{($\alpha=0.05$)}.
Multiple comparison testing was controlled with false-discovery rate (FDR) for multiple testing under dependency \cite{benjamini2001control}. 
\RG{We further tested for an interaction between the occurrence of sympathetic responses and the olfactory stimuli through a within-subject two-way ANOVA on the number of EDA epochs grouped by odor condition (i.e., clean-air, n-butanol, banana, isovaleric acid) and the presence/absence of an SMNA response, respectively \gian{($\alpha=0.05$)}.}

\RG{Finally, we tested for a significant effect of contextual odors and sympathetic responses on the amplitude of each ERP component described in Section \ref{sec:EEG_prep} through separate two-way ANOVAs, with the odor condition (i.e., clean-air, n-butanol, banana, isovaleric acid) and the presence/absence of an EDA-driven sympathetic response as within-subject factors} (p-values corrected with FDR, \gian{$\alpha=0.05$}). Post-hoc comparisons were conducted with a t-test, and the resulting p-values were adjusted with the Bonferroni correction \gian{($\alpha=0.05$)}.

\subsection{Effective connectivity analysis with DCM}
\gian{The analysis of effective connectivity was carried out using SPM12
	\cite{ashburner2014spm12}}. 

We used DCM for ERPs \cite{david2006dynamic, kiebel2008dynamic} to investigate the modulatory effect of odors and sympathetic responses on the effective connectivity.
Specifically, DCM models the observed ERPs by combining a physiologically plausible neuronal model of interacting cortical regions, and a spatial forward model that maps the cortical activity to observed EEG data. Each region is described by three interconnected neuronal subpopulations: interneurons, spiny stellate cells, and pyramidal neurons. Regions are coupled to each other through extrinsic connections, that are distinguished into forward, backward, and lateral according to the hierarchical organization of the cortex \cite{felleman1991distributed}. The effect of administered sensory stimuli on neuronal dynamics is accounted for through specific input connections modeling the afferent activity relayed by subcortical structures to the spiny stellate layer of target cortical regions \cite{david2006dynamic, kiebel2008dynamic}. Notably, such inputs are the same for each experimental condition. Accordingly, differences among ERPs due to either contextual or stimulus attributes are explained by modulatory coupling gains on the connection strengths \cite{david2006dynamic, kiebel2008dynamic}. The activity of pyramidal neurons from each region is then projected to the EEG channels through an electromagnetic forward model which accounts for the field spread effects in the head.

\subsubsection{Network specification}
The DCM for ERP framework explains ERP dynamics as arising from a small number of brain regions \cite{penny2018dynamic}. The selection of which brain regions to include in the network for DCM analysis can be made using either prior knowledge from the literature or source reconstruction techniques. Here, we adopted a group source reconstruction approach based on Multiple Sparse Priors (MSP) implemented in SPM12 \cite{lopez2014algorithmic}. Particularly, while MSP has been shown to potentially reduce localization error with high-density EEG systems \cite{lopez2014algorithmic}, group-inversion yields better results compared to individual inversions by introducing additional constraints on the set of sources explaining subjects' data \cite{litvak2008electromagnetic}. \gian{Operationally, we used channels' position co-registered to the MNI standard head template as provided by SPM.}
\gian{Then, we} inverted ERPs activity on a cortical mesh comprising 8196 dipoles in the time range from -200 to 400 ms. 
\gian{For each subject, we then created contrasts of log power differences in the 0-400 ms time range, collapsed over the experimental conditions, against the prestimulus window (i.e., -200 to 0 ms). These contrast images were smoothed with an 8mm Gaussian kernel to create a 3D volumetric NIFTI image. Images were entered into a one-sample t-test design in SPM12 and we tested for significant changes in power with respect to the prestimulus through an F-contrast.}
\gian{Significant regions (p$<$0.05; Family-Wise Error Rate corrected at the cluster level)} were labeled according with the Automated Anatomical Labeling (AAL) atlas \cite{rolls2020automated}.

\subsubsection{DCM subject-level connectivity analysis}
We performed DCM analysis on the subject-average ERPs relative to the odor conditions, and with and without a sympathetic response. Accordingly, for each subject, we specified a \fb{factor} analysis on the single-subject connectivity, with \textit{Odor} and \textit{Symp} as between-trial factors.
We reduced the data to the first \rhoRev{four principal components (PC) or modes} of EEG channels' mixture. Such a choice is a trade-off between the computational cost of DCM model inversion and the percentage of variance explained by the data \cite{kiebel2006dynamic, litvak2011eeg, rho2022valence}. Notably, reducing ERP data to their first four PC is indicated as sufficient to capture the components of interest \cite{david2006dynamic}. 
We focused model inversion on the 0-400 ms time window with respect to the stimulus onset, through a Hanning window. Finally, we adopted the ERP \gian{neural mass model (NMM)} to model temporal dynamics within and between the network sources.

Concerning the forward model, we modeled the spatial activity of brain sources as equivalent current dipoles (ECD option in SPM12) on the cortex. The passive volume conduction effects on the dipoles' electric field were modeled through a BEM model of the head made of three layers, i.e., cortex, skull and brain, whose conductances were set to 0.33, 0.0042 and 0.33 S/m, respectively.
\\
We allowed the effect of both \textit{Odor} and \textit{Symp} to modulate all the connections of the network, including self-inhibitory effects on each node. 

\subsubsection{PEB group-level connectivity analysis}
We inferred the significant modulatory effects of \textit{Odor} and \textit{Symp} on the group-level connectivity through the PEB framework \cite{zeidman_guide_2019, friston2015empirical}.
Such a framework allows to build statistical hierarchical models where single-subject DCM parameters of interest can be treated as random effects on the group-level connectivity:

\begin{align}
	\label{eq:PEB_singleSubject}
	Y_i &= \Gamma(\theta_i^{(1)}) + \epsilon_i^{(1)} \\
	\label{eq:PEB_group}
	\theta^{(1)} &= (X_b \otimes X_w)\theta^{(2)} + \epsilon^{(2)} 
\end{align}

More specifically, single-subject ERPs $Y_i$ are explained by a DCM $\Gamma(.)$ with unknown parameters $\theta_i^{(1)}$, plus a zero-mean white Gaussian noise residual $\epsilon^{(1)}$. Parameter estimates of interest are then grouped together across subjects (i.e., $\theta_i^{(1)}$) and modeled at the group-level (\ref{eq:PEB_group}) through a General Linear Model (GLM) having design matrix $X = (X_b \otimes X_w)$ and group parameters $\theta^{(2)}$. The $X_b$ matrix models the between-subject effects, whereas the $X_w$ matrix models which single-subject parameters are influenced by
such effects. The $\otimes$ symbol denotes the Kronecker product.
Any unexplained between-subject difference (e.g., non-modeled sources of variations, random effects) is modeled by the zero-mean white Gaussian residuals $\epsilon^{(2)}$. 

We grouped together the single-subject estimates associated with the modulatory effect of \textit{Odor} and \textit{Symp} such that $\theta^{(1)} = [\theta_{Odor}^{(1)}, \, \theta_{Symp}^{(1)}]^T$, and we fitted a PEB model with between-subject design matrix $X_b=1^T$ and within-subject design matrix $X_w = I$ to estimate the average effect of such conditions on each connection of the group connectivity. 

We then applied a greedy search to infer the best combination of group-level parameters $\theta^{(2)}$ describing the average effects of \textit{Odor} and \textit{Symp}. Specifically, we iteratively applied Bayesian Model Reduction (BMR) to obtain an estimate of nested PEB models with/without a particular set of connections, as well as their \RG{Posterior Probability (Pp)} of being the best model describing the observed data \cite{friston2016bayesian}.
We then computed a Bayesian Model Average (BMA) over the models resulted from the last iteration of greedy-search procedure. BMA performs an average of the parameter posterior densities across models, weighted for their Pp. Accordingly, we obtained a set of group-level parameters whose value is no longer conditional on the particular model assumed. Finally, we thresholded the BMA results pruning away those parameters having a Pp of being modulated by either \textit{Odor} or \textit{Symp} lower than 0.95. We based such thresholding on the free-energy of the models estimated during the BMR procedure (see Appendix 3 of \cite{zeidman_guide_2019}).

\section{Results}

\subsection{SAM statistical analysis results}
\begin{figure*}[h]
	\centerline{\includegraphics[width=0.9\linewidth]{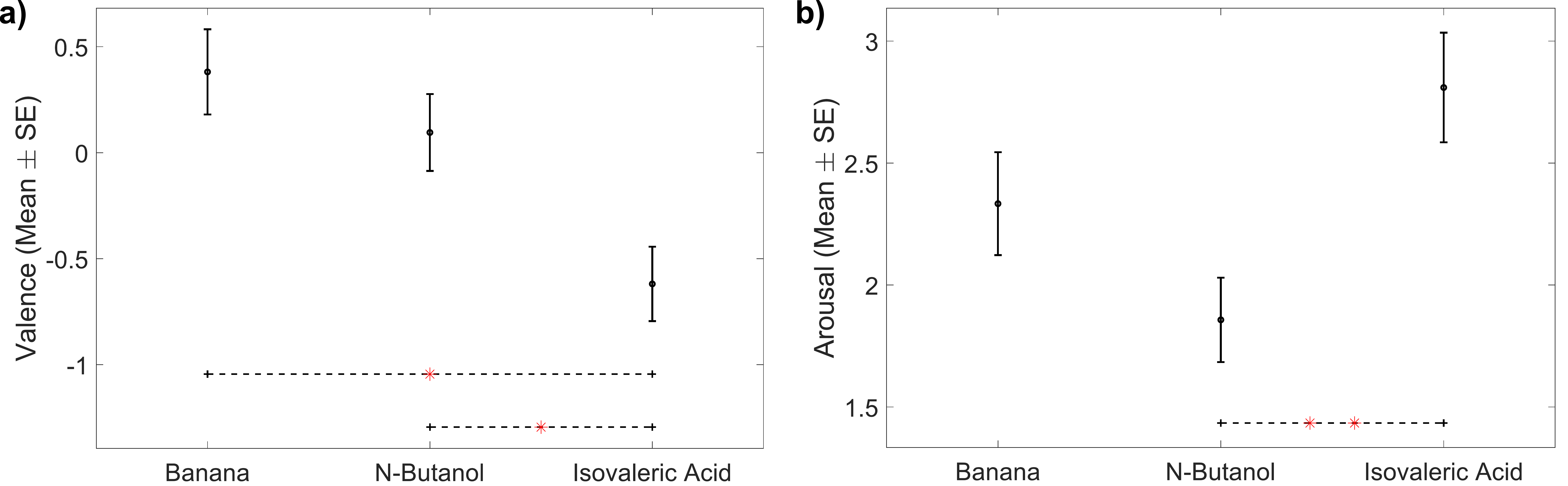}} 
	\caption{SAM statistical analysis results for the a) valence and b) arousal ratings (\rhoRev{Mean} $\pm$ Standard Error (SE)) of the administered odorants (i.e., Banana, Butanol, Isovaleric Acid). Subjects perceived isovaleric acid as significantly more unpleasant ($-0.62 \pm 0.18$) and more arousing ($2.81 \pm 0.22$) than the other odorants.}
	\label{fig:EEG_EDA_SAM_odori}
\end{figure*}

In Fig.\ref{fig:EEG_EDA_SAM_odori} we report the statistical analysis results of the SAM on the administered odorants (\rhoRev{Mean} $\pm$ Standard Error (SE)). Subjects rated isovaleric acid as significantly more unpleasant (valence = $-0.62 \pm 0.18$; Fig.\ref{fig:EEG_EDA_SAM_odori}a) and more arousing (arousal = $2.81 \pm 0.22$; Fig.\ref{fig:EEG_EDA_SAM_odori}b) compared to banana and n-butanol. We did not find any significant difference\fb{s} between the valence and arousal ratings of banana (valence = $0.38 \pm 0.20$; arousal = $2.33 \pm 0.21$) and n-butanol (valence = $0.10 \pm 0.18$; arousal = $1.86 \pm 0.17$). 

On the other hand, we did not find any significant difference for both the valence (clean air = $-0.05 \pm 0.05$; banana = $-0.14 \pm 0.0.06$; n-butanol = $-0.03 \pm 0.05$; isovaleric acid = $-0.09 \pm 0.07$) and arousal (clean air = $2.17 \pm 0.16$; banana = $2.23 \pm 0.17$; n-butanol = $2.18 \pm 0.16$; isovaleric acid = $2.15 \pm 0.17$) ratings of faces among different odor conditions.

\subsection{EDA statistical analysis results}
We did not find any significant difference\fb{s} among odor conditions neither for the average amplitude nor the latency of SMNA responses.
Conversely, 
\gian{we found a significant effect of nSymp following the 2x4 ANOVA} ($F_{6,64}=5.76, p<10^{-5}$; FDR-corrected). More specifically, we found a lower number of stimuli associated with a sympathetic response \rhoEdit{(11.76 $\pm$ 3.06)}, compared to the stimuli without the manifestation of a sympathetic response \rhoEdit{(19.61 $\pm$ 2.63)}, irrespective of the odor condition.

\subsection{ERP statistical analysis results}
\begin{figure*}[h]
	\centerline{\includegraphics[width=0.8\linewidth]{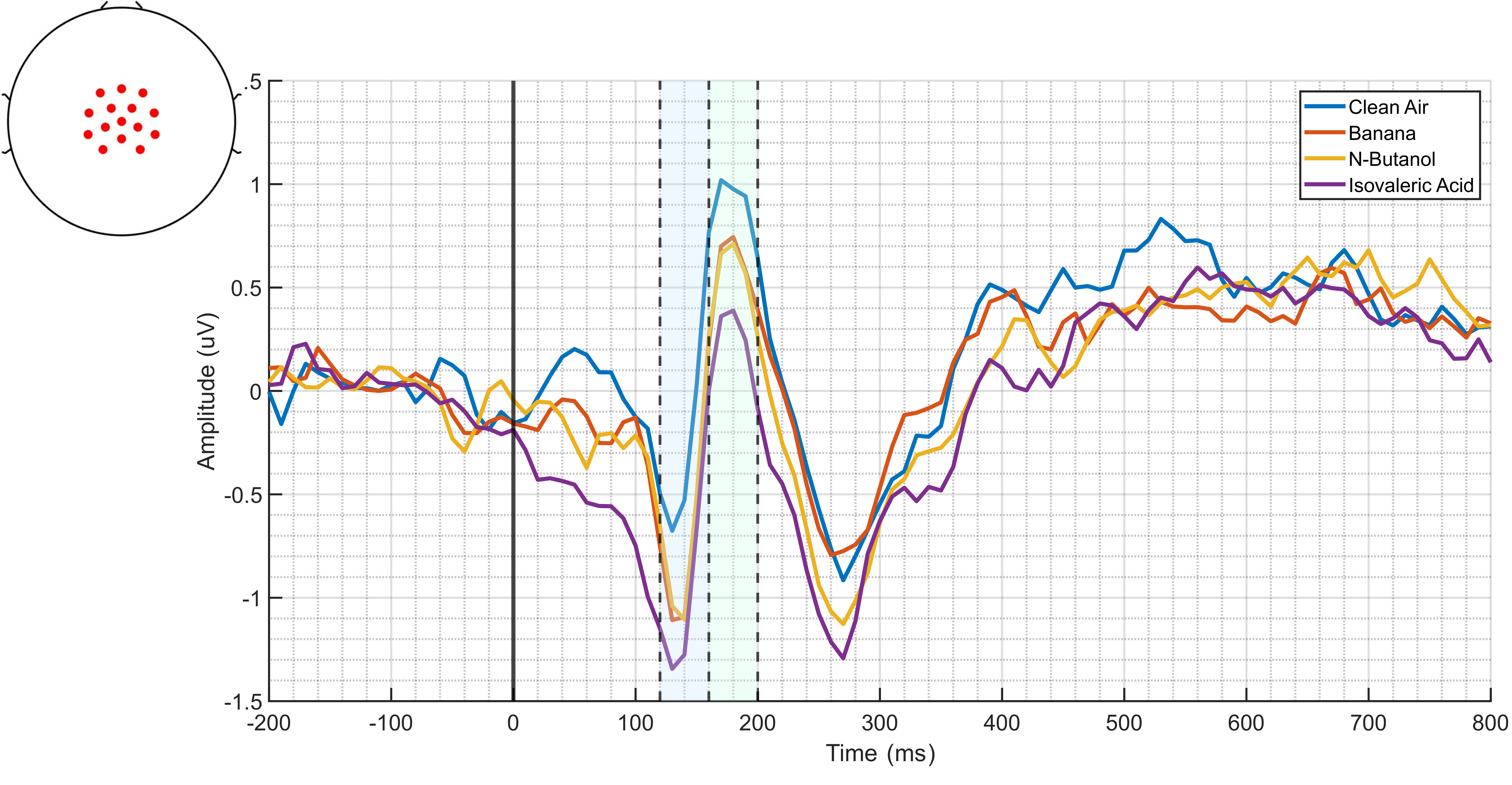}} 
	\caption{ERP grand-averages for the Clean Air (blue), Banana (red), N-Butanol (yellow), and Isovaleric Acid (purple) conditions evaluated at the central ROI, in the time range from -200ms to 800ms with respect to the stimulus onset (i.e., 0ms). ERPs were averaged across 15 electrodes around Cz, schematically represented by the red dots on the scalp map on the top-left of the figure. The blue area highlights a significant effect for the odors on the average N1 amplitude, computed \gian{over} the 120-160 ms interval ($p<0.05$, FDR-corrected). On the other hand, the green area highlights a significant effect for the odors on the average VPP amplitude, computed \gian{over} the 160-200 ms interval ($p<0.05$, FDR-corrected). Post-hoc analysis showed a greater N1 amplitude during Isovaleric Acid, with respect to Clean Air ($p<0.05$, corrected with Bonferroni). Moreover, the VPP was significantly lower for both the Isovaleric Acid and the N-Butanol, with respect to Clean Air ($p<0.05$, corrected with Bonferroni).}
	\label{fig:EEG_EDA_1x4ANOVA}
\end{figure*}

We found a significant effect for the odor condition on the N1  ($F_{6,64}=6.65, p<10^{-4}$; FDR-corrected) and the VPP  ($F_{6,64}=3.63, p<10^{-4}$; FDR-corrected) components at the central ROI around the Cz channel. In Fig.\ref{fig:EEG_EDA_1x4ANOVA}, we report the grand-average ERP for each of the odor conditions,  together with the N1 and VPP significant intervals highlighted in blue and green, respectively, and a schematic representation of the ROI on the scalp map. 
Particularly, post-hoc analysis highlighted a greater N1 amplitude during the administration of isovaleric acid, with respect to clean air ($p<10^{-3}$).
Moreover, we observed a lower VPP amplitude during the administration of both isovaleric acid ($p<10^{-3}$) and n-butanol ($p<10^{-3}$), with respect to clean air.  

\begin{figure*}[h]
	\centerline{\includegraphics[width=0.8\linewidth]{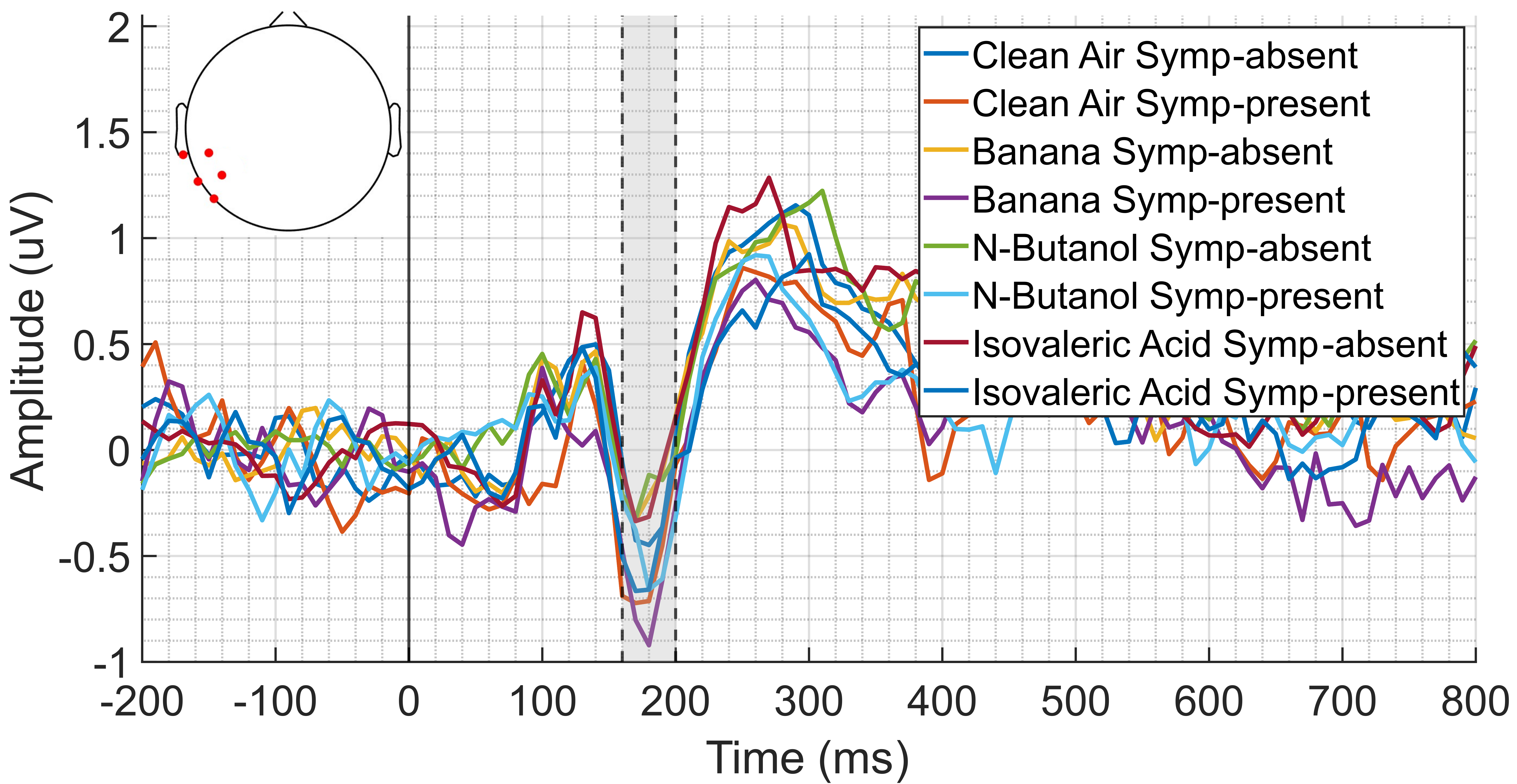}} 
	\caption{ERP grand-averages for the Clean Air Symp-absent (blue), Clean Air Symp-present (red), Banana Symp-absent (yellow), Banana Symp-present (purple), N-Butanol Symp-absent (green), N-Butanol Symp-present (light blue), Isovaleric Acid Symp-absent (brown), and Isovaleric Acid Symp-present (dark blue) conditions,  evaluated at the left ROI around P7, in the time range from -200ms to 800ms with respect to the stimulus onset (i.e., 0ms). ERPs were averaged across 5 electrodes around P7, schematically represented by the red dots on the scalp map on the top-left of the figure. The gray area highlights a significant effect for the \textit{Symp} condition on the average N170 amplitude, computed \gian{over} the 160-200ms interval ($p<0.05$, corrected with Bonferroni). Specifically, the N170 deflection was greater during sympathetic responses (i.e., Symp-present), irrespective of the odor condition.}
	\label{fig:EEG_EDA_2x2ANOVA_Symp}
\end{figure*}

Moreover, we observed a significant effect for the \textit{Symp} condition on the amplitude of the left N170 component ($F_{1,17}=5.10, p=0.03$). Particularly, as reported in Fig.\ref{fig:EEG_EDA_2x2ANOVA_Symp}, the presence of a sympathetic response resulted in the increase of the N170 amplitude. We did not find a significant interaction between \textit{Odor} and \textit{Symp} for any of the ERP components investigated.   

\subsection{Network identification results}
We found activation of the right Inferior Occipital Gyrus (rIOG; MNI coordinates: 48 -76 -2), right Fusiform Gyrus (rFFG; MNI coordinates: 38 -18 -32), right and left Inferior Temporal Gyrus (rITG; MNI coordinates: 52 -38 -20; lITG; MNI coordinates: -50 -38 -22), right and left Medial Temporal Gyrus (rMTG; MNI coordinates: 52 -62 14; lMTG; MNI coordinates: -54 -64 10), and right Secondary Visual Cortex (rVII; MNI coordinates: -10 -98 -10) (p$<$0.05; Family-Wise Error Rate corrected at the cluster level). 

\begin{figure}[h]
	\centerline{\includegraphics[width=1\linewidth]{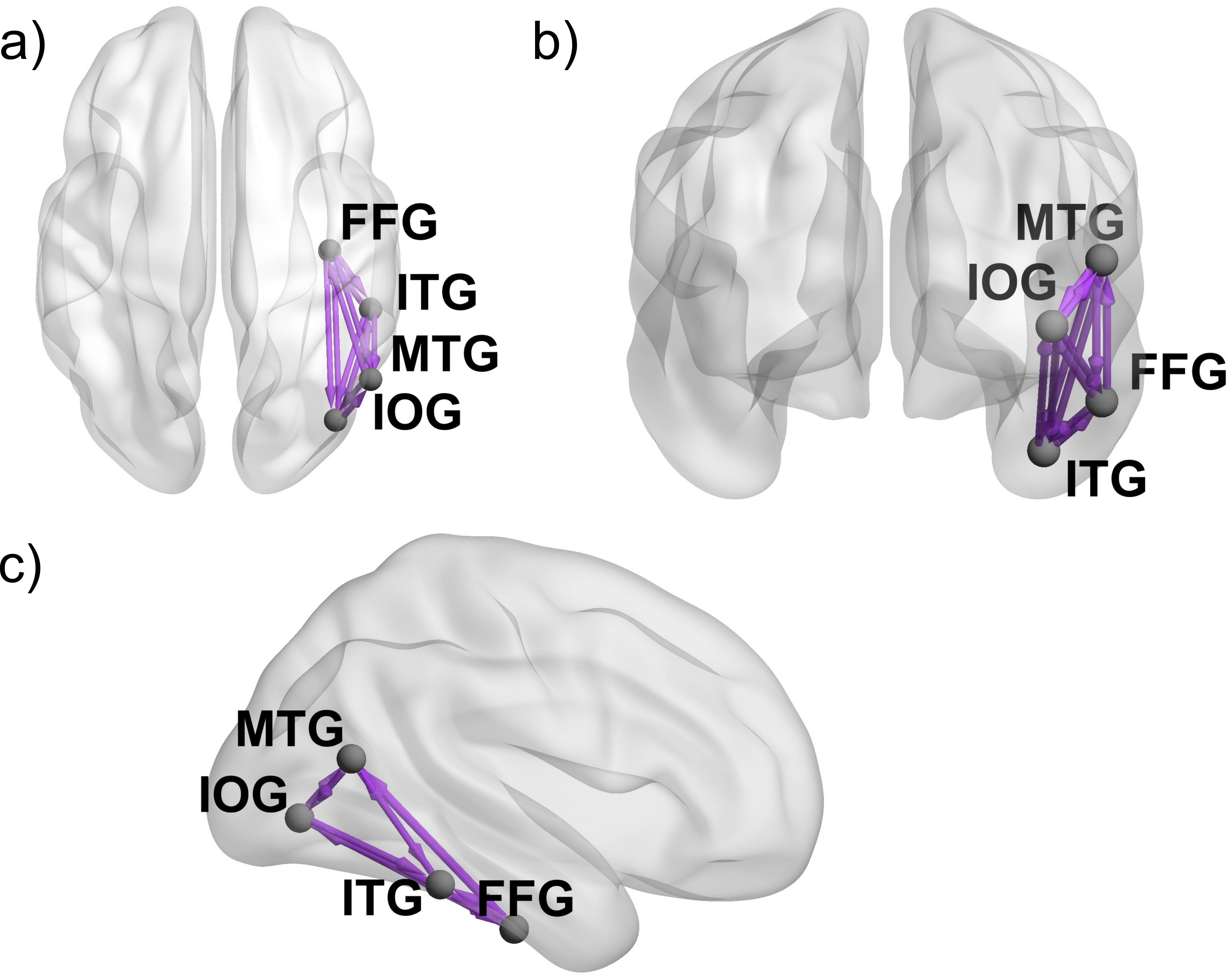}} 
	\caption{Axial (a), coronal (b) and sagittal (c) views of the network to be modeled with DCM. The network includes the Inferior Occipital Gyrus (IOG), Fusiform Face Gyrus (FFG), Inferior Temporal Gyrus (ITG), and Medial Temporal Gyrus (MTG). These nodes were found through a group-inversion based on the MSP approach in SPM12. Nodes were labeled according with the Automated Anatomical Labeling (AAL) atlas. We specified rIOG$\rightarrow$rFFG, rFFG$\rightarrow$rITG and rITG$\rightarrow$rMTG as forward-connections, and rFFG$\rightarrow$rIOG, rITG$\rightarrow$rFFG and rMTG$\rightarrow$rITG as backward connections.}

	\label{fig:EEG_EDA_DCM_network}
\end{figure}

We specified the network for DCM analysis focusing on the IOG, FFG, ITG, and MTG for their central role in the processing of faces and their role in the integration of visual and olfactory stimuli \cite{kessler2021revisiting, li2010effective, elbich2019evaluating, cecchetto2019collect, rossion2003network, elfgren2006fmri}.
Furthermore, in line with the previous literature, we focused on the right visual stream of face processing \cite{kessler2021revisiting, jacques2019inferior, elbich2019evaluating, nguyen2014fusing, botzel1995scalp}. We specified rIOG$\rightarrow$rFFG, rFFG$\rightarrow$rITG and rITG$\rightarrow$rMTG as forward-connections, and rFFG$\rightarrow$rIOG, rITG$\rightarrow$rFFG and rMTG$\rightarrow$rITG as backward connections according with previously reported evidence on the nodes' hierarchical structure \cite{kessler2021revisiting, haxby2000distributed, gobbini2007neural, rolls1992neurophysiological, nguyen2014fusing} (Fig.\ref{fig:EEG_EDA_DCM_network}). 
We hypothesized IOG as the first stage responsible for face processing in our network \cite{uono2017time, li2010effective, jamieson2021differential, nguyen2014fusing}. Accordingly, we modeled the effect of the thalamic sensory input relay of "faces" directly entering the IOG (i.e., the network input).

\subsection{DCM subject-level connectivity analysis}
Following the results of SAM and ERP analysis, we decided to focus our connectivity study on the effects of isovaleric acid and \gian{arousal} enhancement on the visual processing of faces. Accordingly, we built a 2x2 factorial DCM analysis with factors \textit{Odor} (i.e., clean air vs. isovaleric acid) and \textit{Symp} (i.e., sympathetic vs. no sympathetic response).
For each subject, the DCM model was successfully fitted to the observed data without observing any problem of early convergence. The models showed a good representation of the data, with an average explained variance of $90.77\%$.

\begin{figure}[h]
	\includegraphics[width=\linewidth]{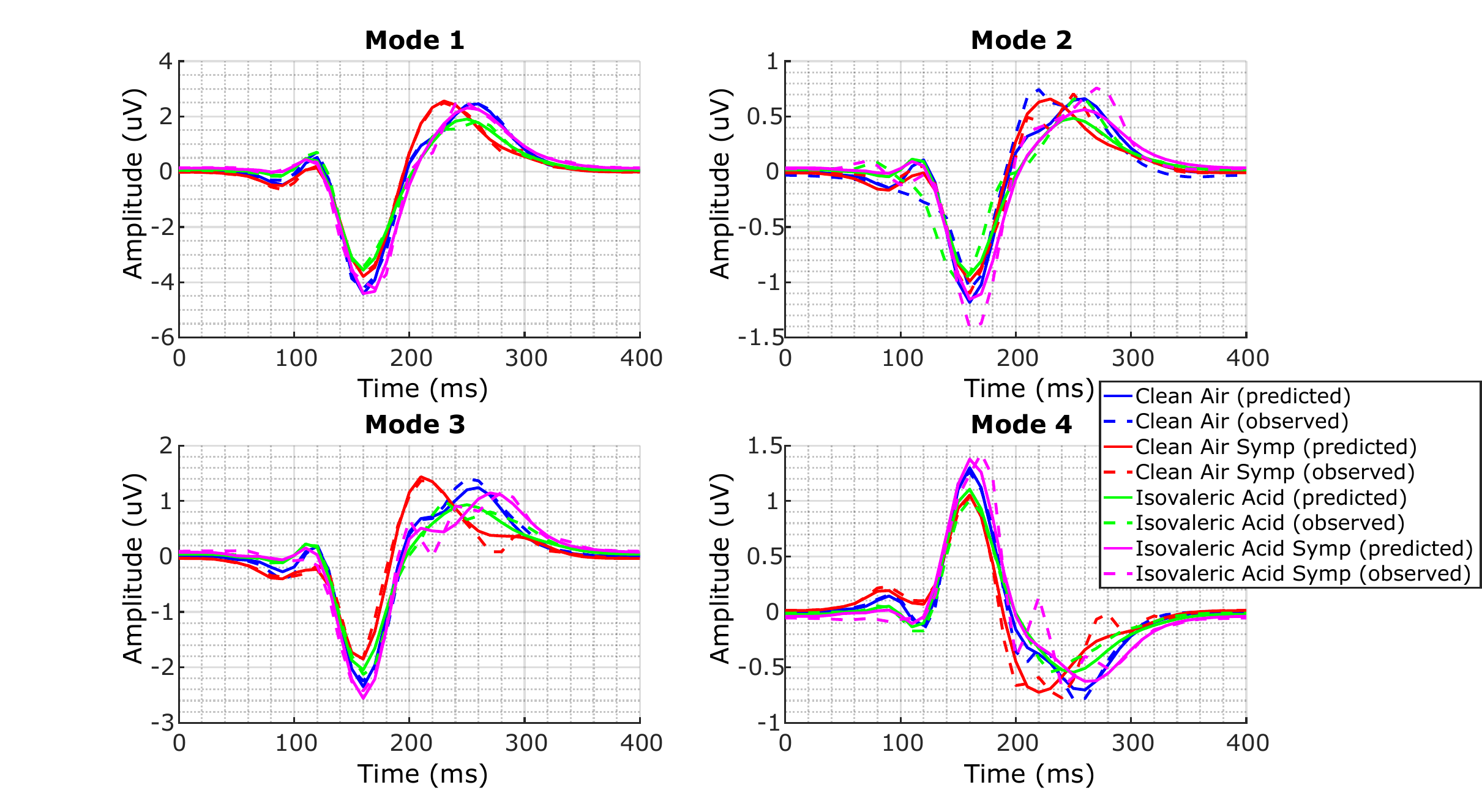}
	\caption{First four principal modes of channels' ERPs for the CleanAir (blue), CleanAir Symp (red), Isovaleric Acid (green), Isovaleric Acid Symp (pink) conditions for an exemplary subject. Dashed lines indicate the observed ERPs, whereas solid lines indicate the ERPs generated by the DCM model. The model was inverted in the 0-400 ms interval with respect to the stimulus onset.}
	\label{fig:EEG_EDA_DCM_firstLevel_exemplarySubject}
\end{figure}

In Fig.\ref{fig:EEG_EDA_DCM_firstLevel_exemplarySubject} we report the result of the ERP reduction to the first four principal modes of EEG channels' mixture, together with the modes predicted by the model, for an exemplary subject. As depicted, PCA channels' reduction yielded a parsimonious yet efficient representation of the data. Specifically, the first three EEG modes represented data variance associated with the P1, N170, and P2 components, whereas the fourth mode represented data variance mainly associated with the VPP. Such a procedure allowed us to focus parameters' estimation on the main features of the observed ERPs, while dramatically reducing noise and computational complexity. 

\subsection{PEB group-level connectivity analysis}
\begin{figure*}[h]
	\centerline{\includegraphics[width=0.8\linewidth]{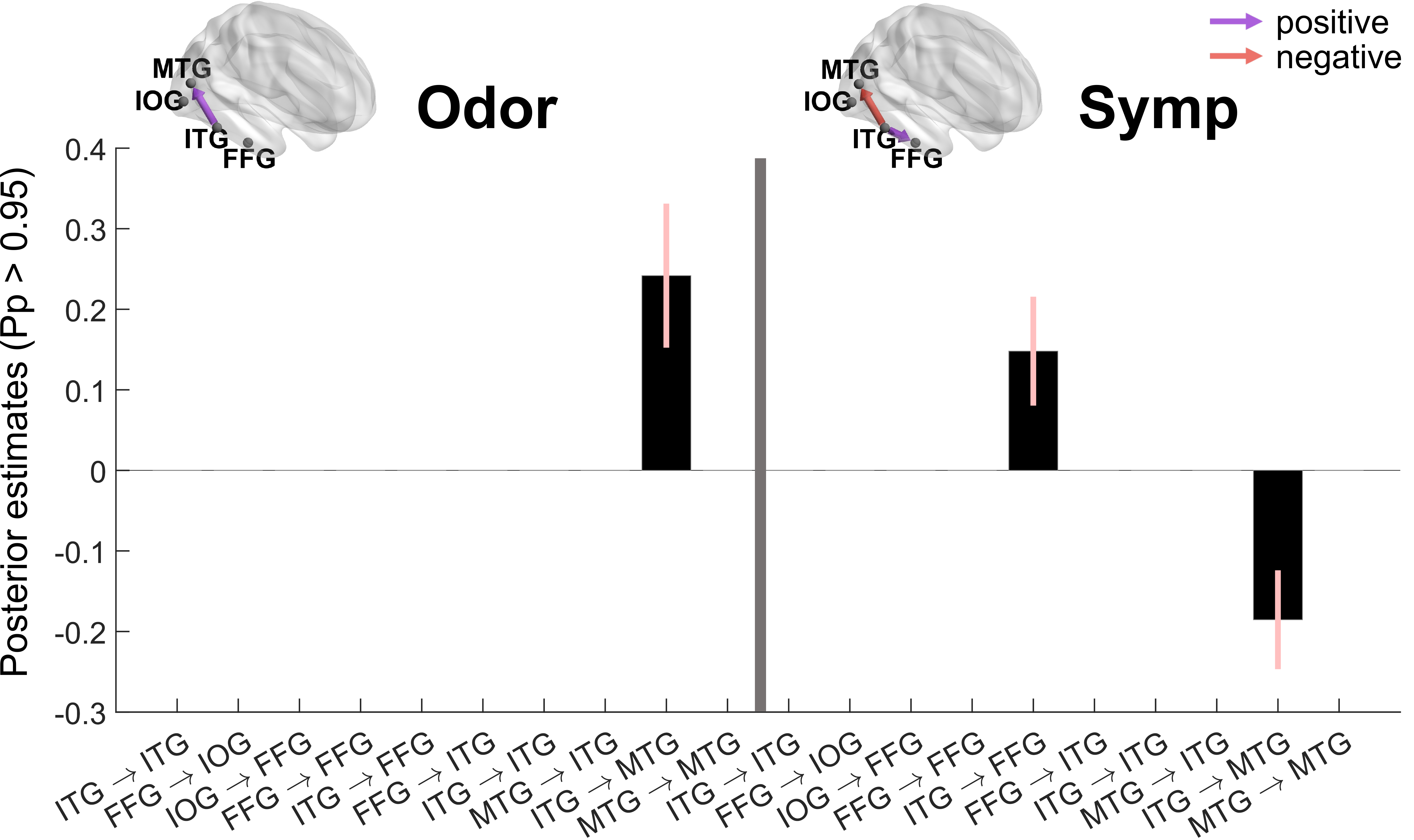}} 
	\caption{Results of BMA on the group effective connectivity among the ITG, FFG, ITG, and MTG. For each experimental condition (i.e., \textit{Odor}, \textit{Symp}), we report the effect sizes associated with their average modulatory effects on the connections' strength. In particular, we report only those parameters having a posterior probability (Pp) $>$ 0.95 of being different from zero. For each significant parameter, the height of the black bar indicates the estimated average posterior effect size, whereas the pink error bar indicates the corresponding 90$\%$ credibility interval. The brain figures depict the rendering of the network on the axial plane. For each condition, connections reporting a significant positive modulation are depicted in purple, whereas connections reporting a significant negative modulation are depicted in red.}
	\label{fig:EEG_EDA_BMA_results}
\end{figure*}

In Fig.\ref{fig:EEG_EDA_BMA_results}, we report the results of the greedy-search and BMA procedure on the PEB parameters associated with the average modulatory effects of the \textit{Odor} (i.e., clean-air vs. isovaleric acid) and \textit{Symp} (i.e., absence vs. presence of a sympathetic response) conditions on the group connectivity. 
These BMA parameters indicate the effect size of the average group-modulatory effect associated with the experimental variables of interest in the connectivity.
In particular, we report only those parameters having a Pp$>$0.95 of being different from zero. The parameters surviving such thresholding are represented by a black bar showing the estimated posterior effect size, and a pink error bar representing the corresponding 90$\%$ credibility interval. 

Concerning the \textit{Odor} condition, we observed a\RG{n increase of the forward connection strength from ITG to MTG (}effect size: $0.24\pm0.09$\RG{)}. This indicates that the administration of isovaleric acid induced a significant (i.e., Pp$>$0.95) excitatory effect on the strength of the ITG$\rightarrow$MTG connection at the group level, with respect to the administration of clean-air. On the other hand, we observed a \RG{decrease of the ITG$\rightarrow$MTG connection strength (effect size: $-0.19\pm0.07$)} associated with the \textit{Symp} condition. This, instead, indicates that the presence of a peripheral sympathetic response induced an inhibitory effect on the strength of such connection at the group level. Finally, we also observed a \RG{strengthening effect} of the \textit{Symp} condition on the backward connection from ITG to FFG \RG{(effect size: $0.15\pm0.07$)}, \RG{indicating that peripheral sympathetic responses were associated with an excitatory effect on such connection}.  

\section{Discussion}
In this study, we investigated the modulatory effect of contextual hedonic olfactory stimuli on the visual processing of neutral faces through the analysis of ERP components and effective connectivity. Particularly, we assumed that
enhanced arousal to the perception of faces could play a role on their integration with olfactory stimuli, improving the SNR of the observed responses
and the consequent effect size of the odors \cite{nieuwenhuis2011anatomical, sokolov1963perception}. To this aim, we applied a novel
methodological approach exploiting EDA-driven SMNA to classify visual stimuli into
two cases: i.e., "eliciting" or "not-eliciting" a stimulus-evoked peripheral sympathetic
response. We included the outcome of this procedure as an additional
experimental factor, together with the valence of contextual odors, into a standard analysis
of ERP components. We further modeled observed ERPs through DCM and PEB to test for the presence of specific cortical connections being concomitantly modulated
by enhanced \gian{arousal} and hedonic olfactory stimuli. Our results highlight the role \gian{arousal} plays on the visual processing of human faces and its multimodal integration with olfactory stimuli.

The ERP analysis revealed an effect of EDA-driven arousal on the left N170 component, irrespective of the odor condition. More specifically, the occurrence of a sympathetic response was associated with an enhanced N170 amplitude. Previous studies have found a significant correlation between the
N170 amplitude and the perceived level of arousal associated with faces, irrespective
of their valence, such that arousing stimuli were associated with increased ERP responses
\cite{almeida2016perceived}. The N170 is considered \fb{to be} the earliest marker of higher-order visual processing,
marking the \fb{structural encoding} of a stimulus as a face. In particular, during the
N170 time window, it is hypothesized that multiple cortical processes underlie the fine-grain\fb{ed} categorization of faces based on \fb{configural processing of their features, e.g., eyes, nose, mouth, and the spatial relationship among them}  \cite{rossion2014understanding, maurer2002many}. In this light, we suggest that the higher N170 amplitude observed during sympathetic responses could represent a marker of enhanced \gian{arousal elicited by} salient features of faces.

In addition to the effect of \gian{arousal}, we observed a greater N1 amplitude during the
administration of isovaleric acid, with respect to clean air. Furthermore, we observed a lower VPP amplitude during the administration of both isovaleric acid and n-butanol, again with respect to clean air. These findings are in line with previous studies reporting an effect of hedonic odors on early visual ERP components \cite{leleu2015contextual, steinberg2012rapid, syrjanen2018background}.
The VPP, together with its inverse N170, is known for being particular\fb{ly} sensitive to
faces \cite{batty2003early, damon2021olfaction,trautmann2013perception}, and it has also been implicated in contextual odor effects \cite{leleu2015contextual}.
Hence, our findings on the VPP modulation by both the unpleasant (i.e., isovaleric acid)
and neutral (i.e., n-butanol) odors, are in agreement with previous studies reporting an
overall effect of olfactory stimuli irrespective of their valence \cite{leleu2015contextual}. On the other hand,
greater amplitudes of the N1/P1 component have been associated with the enhanced
allocation of attention towards faces \cite{adolph2013context, damon2021olfaction}. \fbagain{Overall, given the ERPs results, we can state that hedonic odors generate a contextual effect and thus a top-down influence on face processing in different stages.} In this view, we may suggest that the
administration of isovaleric acid had a greater arousing effect on the early processing
of visual stimuli, compared to the other odorants. This is further corroborated by our
findings on the subjective ratings of odors, showing isovaleric acid as being perceived
as significantly more arousing and unpleasant compared to banana and n-butanol. 
Accordingly,
we decided to carry out the remainder of our study focusing on the contrast
between isovaleric acid and clean air.

Using DCM for ERP and PEB, we investigated the concomitant effect of Odor and
Symp on the brain connectivity at the group level. The study revealed interesting
results. On the one hand, we found that isovaleric acid, compared to clean air,
strengthened the forward connection from the ITG to the MTG. \fbagain{In classical models of face processing \cite{haxby2000distributed, kessler2021revisiting, gobbini2007neural}, the MTG and the superior temporal sulcus (STS) are cortical areas specialized in processing changeable facial features (e.g., emotional expressions, gaze, lip movements). The strengthening of the connection towards these areas due to isovaleric acid may represent a crucial aspect of our study. Variable characteristics of the face are fundamental features for intra-species communication and social interaction \cite{allison2000social}. For this reason, an interpretation of this result may consist \fbRev{of} the fact that a negative odor can influence face processing mostly for interaction purposes. A repulsive smell would enhance the ability \fbRev{to interpret} intentions, expressions and mental states of a possible other person (or threatening agent) in a more efficient way, bringing therefore an evolutionary advantage.} On the other hand, the
same connection suffered from an inhibitory effect when a simultaneous peripheral sympathetic
response occurred, irrespective of the odor condition. Moreover, the occurrence
of such sympathetic responses caused an excitatory effect on the backward connection
from ITG to FFG. Face processing network models \cite{haxby2000distributed, gobbini2007neural} highlight the functional dissociation between the fusiform face area (in the FFG) and temporal areas. As stated above, MTG and STS are designated to process changeable facial aspects, while the FFA is crucial to process invariant aspects of the face to access information related to the identity. According to our results, the facial information processing appears to stop at the FFG stage for faces \fbRev{that generate a sympathetic response}, since forward connections towards the MTG are inhibited while backward connections to the FFG are enhanced. Therefore, we can \fbRev{speculate} that the processing of faces \fbRev{that evoke a sympathetic response may be} focused on identity-related information, to the disadvantage of emotional expressions and other changeable features. \fbRev{In other words, some intrinsic facial features create a sympathetic response which, in turn, may result in a top-down enhanced processing
	of identity.} Speculatively, if a face \fbRev{generates a sympathetic response} it is likely that this individual will feel in jeopardy, or more generally anxious. In this situation, fast processing of the face identity may represent an advantage in order to understand whether s/he is in actual danger and, in that case, adopt the appropriate fight or flight strategies connected to higher arousal.
Moreover, these dynamics seem to suggest a crucial role ITG plays in processing
\gian{sympathetic} arousal that can be associated with odor perception. The ITG has been
described as an associative area integrating information regarding face features from
the FFG and IOG to process the identity of faces \cite{tovee1993neural}. Furthermore, the ITG sends bidirectional projections to limbic areas involved in emotions and memory processing, such as the \rhoEdit{amygdala, hippocampus, entorhinal cortex,} 
and to the frontal cortices \cite{tovee1993neural, van1982parahippocampal}. In this context, previous studies have reported a role for the ITG in visual short-term memory \cite{fuster1990inferotemporal, iwai1990monkey, miyashita1990associative},
as well as in hedonic olfactory tasks and in the perception of faces with background
odor cues \cite{cecchetto2019collect, riley2015altered}. \fbagain{In this light, the ITG appears to be a hub in which dissociation in different pathways for affective processing of faces are handled.}

The methodology proposed in this study could represent an effective means to quantitatively study the effect of sympathetic arousal on both the ERPs and effective connectivity. Indeed, in previous studies, we adopted subjective ratings as a means to quantify arousal and investigate its central correlates at the connectivity level \cite{rho2021odor, rho2022valence}. However, affective ratings may deviate from physiological responses \cite{rho2022valence}, which are supposed instead to represent an objective window on \fb{emotional stimuli} processing (see e.g. \cite{ledoux1998emotional}). Other studies instead investigated the coupling between EDA and EEG dynamics through correlation \cite{mobascher2009fluctuations, stuldreher2020physiological, posada2019phasic, lim1996relationship}, phase and amplitude coupling \cite{kroupi2013phase}, and coherence measures \cite{ural2019wavelet}. Yet, to the best of our knowledge, a methodology to investigate the effect of sympathetic arousal on the EEG effective connectivity as estimated through EDA is missing. In this light, we proposed an approach based on the identification of enhanced sympathetic arousal through the analysis of EDA and the connectivity framework of DCM for ERP. Particularly, we adopt\fb{ed} cvxEDA as an efficient approach to recover an estimate of the hidden SMNA, and identif\fb{ied} the time instants at which stimulus-evoked sympathetic responses have occurred. The DCM framework then allowed to investigate such effects on the effective connectivity with the highest level of physiological plausibility.

It is worth highlighting that the observed peripheral sympathetic responses could be attributed either to the effect of background olfactory stimuli, the presentation of visual faces, or the combination of both. With this regard, our EDA analysis results did not show any significant effect\fb{s} of the administered background olfactory stimuli on either the amplitude, the latency, or the frequency of the sympathetic responses following the presentation of human faces. Accordingly, it is reasonable to assume that the observed peripheral sympathetic responses time-locked to the presentation of faces could be associated with the intrinsic arousing properties of the faces themselves. Thus, the proposed methodology allows accounting for two distinct, yet concomitant, effects on the EEG cortical connectivity, i.e., the modulation of contextual hedonic odors on the perception of human faces, and the modulation of enhanced \gian{sympathetic arousal elicited by} the visual presentation of faces.
Nevertheless, to the best of our knowledge, this is the first investigation of EDA dynamics in the scenario of the multimodal integration between visual faces and olfactory stimuli. Hence, further investigations will be necessary to corroborate this hypothesis.

The choice of focusing our DCM analysis only on the right hemisphere was supported by the literature\cite{kessler2021revisiting, jacques2019inferior, elbich2019evaluating, nguyen2014fusing, botzel1995scalp, walla2008olfaction} and allowed us to limit the number of parameters (i.e., nodes, connections) at the advantage of less uncertainty on their estimates. However, this can be seen in contrast to the ERP analysis \fb{that} revealed an effect of sympathetic arousal on the left N170 component, while no effects were observed in the right hemisphere. Nevertheless, it is worth noting that multiple sources underlie the processing of faces in the N170 (160-200)ms interval. Hence, ERP components may reflect the global activation of the face perception system, potentially leading to small or null effects \cite{rossion2014understanding}. 
In this view, DCM offers a powerful means to investigate the effect of sympathetic arousal to a deeper extent through a physiologically plausible description of how observed dynamics are associated with its underlying cortical underpinnings. 
Accordingly, future works should increase statistical power by including more subjects and extend our network to also include left-hemisphere ITG and MTG found during ERP inversion. 

The framework proposed does not include the amplitude of the SMNA in the model. This is a limitation of the model since we cannot exclude that a potential relationship between central responses and the magnitude of the sympathetic neural bursts could exist. Yet, modeling such an effect through DCM would require to specify a single-trial within-subject modulatory input where each administered stimulus is associated to its respective level of sympathetic arousal (as quantified through the amplitude of SMNA). To the best of our knowledge, the DCM for ERP framework is limited to model between-trials effects (i.e., differences among grand-average ERPs), rather than single-trial effects. Accordingly, we limited our methodology to the evaluation of connectivity differences in the binary case of the presence/absence of an evoked peripheral sympathetic response, without taking into account their magnitude.

SMNA responses may be triggered by a number of factors different from the perception of the designed administered stimulus (in this context, human faces). Particularly, spontaneous fluctuations, i.e., sudomotor nerve responses which stem from uncontrolled cognitive and emotional processes \cite{boucsein2012electrodermal}, may lead to the erroneous detection of a face-evoked peripheral sympathetic responses. While we attempted at mitigating such effects constraining stimulus-evoked SMNA bursts in the 1-5s time window after stimulus onset \cite{boucsein2012electrodermal}, we cannot exclude that errors in the binary classification of the stimuli may still occur.

Another potential limitation of our methodology concerns the relationship between EDA-driven peripheral sympathetic responses and EEG central responses. Specifically, we could not explicitly include EDA dynamics into the analysis of effective connectivity, as the DCM framework only allows to model modulatory effects on the observed EEG data due to either contextual factors or properties of the administered stimuli \cite{kiebel2008dynamic}. Accordingly, nothing can be concluded about the causal relationship between EEG and EDA responses. To overcome these limitations, future studies will aim develop a framework including EDA dynamics as a representative node in the connectivity network.

\section{Conclusion}
In this work, we proposed a novel methodological approach based on the analysis of EDA and EEG to investigate the effect of contextual hedonic odors and \gian{sympathetic arousal} on the visual processing of neutral faces. Our main findings highlighted a higher N1 component during the unpleasant odor condition, whereas enhanced \gian{arousal} to faces increased the N170 component. Moreover, both factors exerted a significant modulatory effect on the effective connectivity among cortical areas involved in face processing. Particularly, the unpleasant odor strengthened the forward connection from ITG to MTG, whereas the same connection was inhibited by the simultaneous occurrence of \gian{face-evoked sympathetic responses}. \fbagain{These results may suggest that, on the one hand, face processing in the context of an unpleasant odor are more focused on changeable facial aspects related to social interaction. On the other hand, increased \gian{arousal} appears to enhance identity processing in the FFG.}  
Future works will include SMNA dynamics as a representative node in the connectivity network, to investigate EDA-EEG causal interactions and deepen the understanding of \gian{sympathetic arousal's} role on the visual-odor multimodal integration.

\section*{Conflict of interest statement}
The authors declare no conflict of interest. 

\ack
The research leading to these results has received partial funding from the Italian Ministry of Education and Research (MIUR) in the framework of the CrossLab project (Departments of Excellence). This research has received partial funding from European Union Horizon 2020 Programme under grant agreement n 824153 of the project “POTION: Promoting Social Interaction through Emotional Body Odours”. Research partly funded by PNRR - M4C2 - Investimento 1.3, Partenariato Esteso PE00000013 - “FAIR - Future Artificial Intelligence Research” - Spoke 1 “Human-centered AI”, funded by the European Commission under the NextGeneration EU programme. This publication was produced with the co-funding European Union - Next Generation EU, in the context of The National Recovery and Resilience Plan, Investment 1.5 Ecosystems of Innovation, Project Tuscany Health Ecosystem (THE), Spoke 3 “Advanced technologies, methods, materials and heath analytics” CUP: I53C22000780001

\normalsize

\section*{References}
\bibliography{bibliography}

\begin{thebibliography}{100}
\bibitem{rossion2003network}
A network of occipito-temporal face-sensitive areas besides the right middle
  fusiform gyrus is necessary for normal face processing.
\end{thebibliography}
\bibliographystyle{vancouver}

\end{document}